\documentclass[conference, 10pt]{IEEEtran}
\IEEEoverridecommandlockouts
\usepackage{cite}
\usepackage{amsmath,amssymb,amsfonts}
\usepackage{algorithmic}
\usepackage{graphicx}
\usepackage{textcomp}
\usepackage{xcolor}
\usepackage{cleveref}
\usepackage{siunitx}

\usepackage{comment}

\usepackage{tabularx,booktabs}

\usepackage{booktabs,ragged2e}

\usepackage[flushleft]{threeparttable}

\usepackage{subcaption}

\newcommand\philipp[1]{\color{blue}\textbf{(Philipp}\color{black}}

\usepackage[numbers]{natbib}

\def\BibTeX{{\rm B\kern-.05em{\sc i\kern-.025em b}\kern-.08em
    T\kern-.1667em\lower.7ex\hbox{E}\kern-.125emX}}
\begin{document}
\bstctlcite{IEEEexample:BSTcontrol}

\title{On Comparing and Enhancing Common Approaches\\ to Network Community Detection}

\author{\IEEEauthorblockN{Niko Motschnig, Alexander Ramharter, Oliver Schweiger, Philipp Zabka, Klaus-Tycho Foerster}\thanks{An abbreviated version of this article appears in~\cite{globecom}.}
\IEEEauthorblockA{\textit{Faculty of Computer Science},
\textit{University of Vienna},
Vienna, Austria \\
\{nmotschnig, aramharter, oschweiger, pzabka, ktfoerster\}@cs.univie.ac.at}
}

\maketitle
\thispagestyle{plain}
\pagestyle{plain}
\begin{abstract}
In this work, we explore four common algorithms for community detection in networks, namely Agglomerative Hierarchical Clustering, Divisive Hierarchical Clustering (Girvan-Newman), Fastgreedy and the Louvain Method. 
We investigate their mechanics and compare their differences in terms of implementation and results of the clustering behavior on a standard dataset. 
We further propose some enhancements to these algorithms that show promising results in our evaluations, such as \textbf{self-neighboring} for Neighbor Matrix constructions, a deterministic slightly faster version of the Louvain Method that favors less bigger clusters and various implementation changes to the Fastgreedy algorithm.
\end{abstract}





\maketitle

\section{Introduction}
\subsection{Motivation}
Community detection is a hot topic in network research, with a wide field of possible applications. 
As networks and graphs are able to offer abstract representations of many different domains, such as social dynamics, disease spread, customer behaviour, and many more, community detection can be employed to analyze and solve a whole range of different problems especially in the field of sociology. 
It is also applicable in other areas, such as, e.g., biology (protein-protein interaction networks), computer science (for instance finding groups of sites dealing with similar topics), and disease control, e.g., in cattle-trade networks \cite{brzoska2020hierarchical}.

Whereas there are a multitude of \emph{new} proposals for community detection algorithms~\cite{FortunatoSanto2010Cdig}, we in this work follow a different research direction.
Instead of designing the next iteration of some algorithmic variant,
we take a step back and investigate the \emph{technical implementation} of common algorithms, based on four selected examples: Agglomerative~\cite[\S4.2]{FortunatoSanto2010Cdig} and Divisive (Girvan-Newman)~\cite{girvanNewman} Hierarchical Clustering, the Louvain Method~\cite{blondel2008fast}, and Fastgreedy~\cite{ClausetAaron2004Fcsi}, which will be presented in the following sections.

To this end, we illustrate and study the general workings of each of these algorithms, give details on their implementation, and discuss on how they as thus differ in terms of results and clustering behavior.
In this investigation, our aim is on the one hand to provide a different perspective on these algorithms, namely, from a technical point of view, but on the other hand, our approach also allows us to provide ideas for new enhancements and optimizations of these seminal works, also outlining possible future research directions.

To this end, we discuss related work in \S\ref{sec:relatedWork}, present the selected algorithms in \S\ref{sec:selectedAlgorithms}, discuss results and adaptions to them in \S\ref{sec:discussion}, concluding and proposing next steps in \S\ref{sec:conclusion}.


\subsection{Contributions}
We propose, implement from scratch, and evaluate a range of adaptations to these well-known algorithms, namely:

\begin{enumerate}
    \item \textbf{Self-neighboring:} for \textit{agglomerative hierarchical clustering} (using the euclidean distance as the distance function to minimize), \textbf{self-neighboring} changes how the neighbor matrix is being constructed and manages to close "gaps" in the clustering results that would otherwise happen without \textbf{self-neighboring}.
    \item A \textbf{deterministic} \textit{Louvain method} that is slightly faster than its counterpart and creates smaller bigger clusters.
    \item An \textbf{enhancement} and \textbf{simplification} of the \textit{Fastgreedy algorithm} concerning the usage of data structures to be slightly faster and also easier to implement, with the trade-off of more memory usage. 
    \item We \textbf{validate} the proposed adaptation of the \textit{Girvan-Newman} method by Meghanathan on a standard data~set.
    \item Lastly, in order to guarantee reproducibility and facilitate other researchers to build upon our work, we make our \textbf{source code publicly available}~\cite{Repo}.

\end{enumerate}

\section{Background}
\label{sec:relatedWork}

There are various methods to perform community detection or clustering in (social) networks.
One of the most well-known overviews of this area is by Fortunato~\cite{FortunatoSanto2010Cdig}\footnote{Nearly \num{10000} citations on Google Scholar.}, who performed an extensive survey on the topic and compared the different algorithms in terms of quality (using normalized mutual information) and runtime complexity.
We follow his work to select four common approaches to community detection, to investigate and compare their implementation details, and to propose and evaluate different enhancements.


A first approach of interest is the \emph{Louvain Method} by Blondel et al.~\cite{blondel2008fast}, a greedy optimization algorithm for community detection.
The Louvain Method uses modularity as its quality metric, showing positive results in comparison with other community detection algorithms on multiple data~sets~\cite{blondel2008fast}.


Another popular method is by means of \emph{hierarchical clustering}.
As there various approaches in this area, we give a brief overview here:
Abbas compares them to k-means, self-organization maps, and expectation maximization based techniques and categorizes the kind of data that fits best for each approach~\cite{comparison1}.
Others compare the performance of different linkage function for agglomerative techniques, such as Yu et al.~\cite{comparison2} evaluating their technique based on single and complete link clustering. 
Further works investigate new metrics and algorithms through which to further optimize hierarchical clustering. 
For instance, Charpentier~\cite{mastermodularity} introduces a new way of doing hierarchical clustering using a modularity score called 'Paris' and Bonald et al.~\cite{nodepair} present another new technique utilizing Node Pair Sampling.
Bateni et al.~\cite{affinity} developed Affinity for Google Research, a hierarchical clustering approach that is able to handle trillions of nodes through MapReduce.
Our work focuses on understanding and visualizing the differences of agglomerative clustering based on fundamental decisions such as distance and linkage functions, and moreover we will provide some intuition on how the clustering algorithm behaves based on the underlying choices and how these can affect the structure of the resulting clusters. 

Furthermore, Girvan et al. \cite{girvanNewman} proposed a method for detecting community structures using a concept called ``Edge Betweenness''. They focused on social and biological networks like the now popular graph based on the \textit{"Karate Club dataset"} by Zachary~\cite{zachary1977information}. The algorithm can be described as a  divisive hierarchical clustering algorithm and is nowadays well-known under the name \emph{Girvan-Newman} algorithm.

Lastly, regarding \emph{modularity based community detection} algorithms, the Louvain Method from Blondel et al.~\cite{blondel2008fast} is a more modern development in this area. 
The work of Clauset et al.~\cite{ClausetAaron2004Fcsi} can be seen as an enhancement of the algorithm from Newman and Girvan \cite{NewmanM.E.J2003Fafd} and also implements greedy optimization techniques based on modularity. 
Moreover, Clauset~\cite{ClausetAaron2004Fcsi} performed some shortcuts regarding the optimizations and included more sophisticated data structures. 

\section{On Investigating Community Detection}
\label{sec:selectedAlgorithms}

In this section, we investigate the four selected standard approaches to community detection, namely, Agglomerative Hierarchical Clustering (\S\ref{sec:AC}), Divisive Hierarchical Clustering (\S\ref{sec:GirvanNewman}), Fastgreedy (\S\ref{sec:FG}), and the Louvain Method (\S\ref{sec:Louvain}).
To this end, for each approach, we first provide an overview, and then present our implementation details and test configurations, which motivate possible enhancements.

\subsection{Agglomerative Hierarchical Clustering}
\label{sec:AC}
\subsubsection{Overview}
\label{sec:ACDesc}
Agglomerative Clustering describes a bottom-up approach: given a network, each node starts off as its own cluster. Step by step, the closest nodes/clusters are joined together. 
This step is repeated until there is only one cluster left, containing all nodes in the network. 
As the word \textit{closest} suggests, there needs to be a way to determine what \textit{close} indicates, performed by a distance function. 
The distance function is always a function between two nodes \textit{i} and \textit{j}, and can vary highly in its complexity. We will test some widely deployed distance functions and see how they change the clustering behaviour later on. 
Additionally, there needs to be a method to not only measure the distance between nodes, but also between clusters, for which we employ \textit{linkage functions}. 
The linkage function describes how to apply the distance function to a \textit{group} of nodes. 
We will discuss how this affects the algorithm later in this section.

Agglomerative Clustering describes a bottom-up approach where nodes or a group of nodes are merged one-by-one until there is only one cluster left.
The deciding elements for such an approach are which distance function and which linkage function to use.
The distance function gives the definition of ``closeness'' within the graph and the linkage function describes how to apply the distance function to node groups.

Other elements that need to be known prior to running the algorithm are the degree of each node \textit{i}, denoted as $k_{i}$, and the Neighbor Matrix $n$, where each element $n_{i,j}$ describes the number of adjacent nodes which \textit{i} and \textit{j} share in common. 
Some distance functions also require to know the overall number of nodes in the network \textit{N}.

Lastly, we use a horizontal separation line (HSL) that describes the strictness of the grouping. 
A dendrogram encodes the order in which the nodes and clusters are joined together in a tree-like structure, so that we can define the HSL to divide the final dendrogram into separate groups again. 
The decision on where to place the HSL has a major influence on the final results, as every vertical dendrogram line the HSL crosses will define its own cluster. 
The higher up the HSL is placed, the fewer the amount of final clusters.


\subsubsection{Implementation and Test Configurations}
\label{sec:ACImpl}
Our implementations~\cite{Repo} are designed to be modular, such that core components can be swapped and replaced according to the use case.
To this end, one can utilize different distance functions, linkage functions, and height of the HSL (among others), and plug in further functions.

We compare three linkage functions and analyzed their behavior, also in comparison to the other algorithms. The distance function used was the network-centric version of the Euclidean distance, defined as
$d_{ij}=k_i + k_j - 2n_{ij}$,
, meaning the distance between two nodes is the sum of their degrees minus the amount of nodes that neighbor both. 
Using this distance function gave results comparable to the other algorithms, as nodes will be grouped first by their network-centric spatial distance.

Concerning the linkage function, there are multiple ways of applying the distance function to determine the distance $d(A,B)$ between two clusters \textit{A}, \textit{B}. We list the most common:




\begin{enumerate}
    \item \emph{minimum} or \emph{single-linkage}: \[d(A,B) = min\{d(a,b) | a \in A,b \in B\}\]
    \item \emph{maximum} or \emph{complete-linkage}: \[d(A,B) = max\{d(a,b) | a \in A,b \in B\}\]
    \item \emph{average linkage}: \[d(A,B) = \frac{1}{|A||B|} \sum_{a \in A}\sum_{b \in B}d(a,b)\]
\end{enumerate}

Finally, the last important parameter to pass in is the HSL height. 
The lower the value, the higher up the HSL will be placed, thus reducing the number of final clusters. 
Our implementation allows to either pass in this value in an absolute or relative level: absolute meaning the precise step-number (counting from the top downward), or a relative level $rel$ where $0 \leq rel \leq 1$, $0$ meaning a line placed at the very top of the dendrogram and $1$ meaning the very bottom.

This set of parameters and functions was the baseline for experimenting with our own implementation of the algorithm and inspecting their differences in terms of resulting clusters. We will discuss these results and further adaptations in \S\ref{AggloDiscuss}.

\subsection{Divisive Hierarchical Clustering (Girvan-Newman)}
\label{sec:GirvanNewman}
\subsubsection{Overview}
\label{sec:GirvanNewmanDesc}
The Girvan-Newman algorithm uses a Divisive Clustering approach. 
In other words, the algorithm deletes edges to form clusters.
It was published in 2002 by Girvan and Newman~\cite{girvanNewman} and describes the concept of ``Edge Betweenness Centrality'' (EBC), defined as the number of shortest paths through an edge.
The Girvan-Newman algorithm uses this value to decide which edge to ``cut''.
It consists of 4 steps:
\begin{enumerate}
\item Calculate the betweenness for all edges in the network,
\item remove the edge with the highest betweenness,
\item recalculate betweennesses for all edges, 
\item and repeat from Step 2 until no edges remain.
\end{enumerate}

This calculation
can be divided into 3 phases:
\begin{enumerate}
    \item Creating a tree using breadth-first search for each node, 
    \item setting the number of shortest paths for each node, 
    \item and calculating the edge betweenness score bottom-up.
\end{enumerate}

Calculating the edge betweenness score for each node in a tree is performed by $e_{ij}=(s_{j}+\sum\limits_{n} e_{ni})/({s_i})$, where
\begin{itemize}
    \item $e_{ij}$ is the edge betweenness between node $i$ and node $j$. $i$ is below $j$ in the tree.
    \item $s_{j}$ describes the number of shortest paths of the node $j$ and $s_{i}$ stands for the number of shortest paths of node~$i$.
    \item $\sum\limits_{n} e_{ni}$ is the sum of all edge betweennesses from the neighbours of node $i$ below it. 
\end{itemize}

After finishing this procedure, the resulting edge betweennesses in the trees are summed up together and build the edge betweennesses of the original graph.
Next, the edge with the highest score is cut and we recalculate the betweennesses for all edges affected by the removal before we repeat this process.

\subsubsection{Implementation}
At the core of the program is the breadth-first search, used for the calculation of the edge betweenness, but also for labeling the communities. 
As the algorithm would continue to cut edges until every node is a community itself, the user needs to specify the maximum number of wanted communities as a parameter.
However, there is no clear answer on how many communities to cluster to, as it varies from network to network.
To obtain comparable results we use the network based on the \textit{Karate Club dataset} from Zachary~\cite{zachary1977information}, which became popular after its' use by Girvan et al. \cite{girvanNewman}.
The results are discussed in \S\ref{sec:girvanNewmanResults}.

\subsection{Louvain Method}
\label{sec:Louvain}
\subsubsection{Overview}
\label{sec:LouvainDesc}
The Louvain Method was published in 2008 by Blondel et al.~\cite{blondel2008fast}. It was named after the University of Louvain in Belgium where the authors worked.

It is a heuristic-based greedy approach that aims to partition the graph into communities that optimize the modularity score, where modularity is a score between -1/2 and 1,\footnote{wrongly reported to be -1 to 1 in the original paper \cite{blondel2008fast}} where the value of 1 can only be achieved on a graph with no edges and the value of -1/2 is only achievable on bipartite graphs if one clusters nodes along their parts~\cite{brandes2007modularity}.
A heuristic is chosen as the actual modularity optimization is NP-complete~\cite{brandes2006maximizing}.
Modularity measures the density of links inside clusters/communities in comparison to links between them, or similarly said, it is a measure for the strength of division of a network into modules. The formula for the computation of the modularity in weighted undirected networks is \cite{FortunatoSanto2010Cdig,blondel2008fast}:
$Q=\frac{1}{2m}\sum_{ij} [A_{i,j} -\frac{k_{i}k_{j}}{2m}]*\delta (c_{i},c_{j})$ , where 
\begin{itemize}
    \item $m$ is the sum of all edge weights in the network (if the weight is always 1 it is just the total number of edges).
    \item $i$ and $j$ are nodes, $c_{i}$ and $c_{j}$ denote the community assignments of the respective node.
    \item $A_{i,j}$ denotes the edge weight between nodes $i$ and $j$.
    \item $k_{i}$ and $k_{j}$ stands for the sum of weights of incoming nodes to $i$ and $j$ respectively.
    \item $\delta (c_{i},c_{j})$ is the Kronecker-delta-function. 
    \end{itemize}

The Louvain method uses this formula indirectly as an objective function that gets optimized in a heuristic based greedy way. The algorithm consists of 2 distinct steps \cite{blondel2008fast}:
\begin{enumerate}
    \item For each node $i$ in the graph calculate the change in modularity that happens when removing it from its current community $c_{i}$ and adding it to every of its neighbouring communities. Then perform the swap that results in the biggest increase in modularity. If no positive change is observed the algorithm terminates.
    \item Merge all nodes belonging to the same community together into one node where all internal connections (from nodes of community $c_{i}$ to nodes of $c_{i}$) become a self-loop to the same node and all external/outgoing connections (from nodes of a community $c_{i}$ to nodes of a different community $c_{j}$) are also merged into one edge. The weight of the merged edge is the sum of edge-weights from all edges merged together.
\end{enumerate}

In order to calculate the change in modularity  when node $i$ gets assigned to the community $c_{j}$ of one of its neighbours in Step 1), a more efficient formula is used that does not require to compute a sum over all nodes~\cite{blondel2008fast}, by setting $\Delta Q=$
$ \left[\frac{\Sigma_{in}+2k_{i,in}}{2m}-\left (  \frac{\Sigma_{tot}+2k_{i}}{2m}\right)^{2}\right] - \left [ \frac{\Sigma_{in}}{2m}-\left(\frac{\Sigma_{tot}}{2m} \right )^{2} - \left ( \frac{k_{i}}{2m} \right )^{2}\right]$, where


\begin{itemize}
    \item $k_{i,in}$ is the sum of weights of all edges from node "$i$" to target Community $c_{j}$.
    \item $k_{i}$ stands for the sum of weights of all edges incoming to node $i$.
    \item $\Sigma_{in}$ denotes sum of weights of internal nodes of target Community $c_{j}$ (sum of weights of edges between nodes of Community $c_{j}$).
    \item $\Sigma_{tot}$  equals the sum of weights of all edges that go to the target Community $c_{j}$ (also includes $\Sigma_{in}$).
\end{itemize}

\noindent \textit{However we believe this not to be the ``complete'' formula to calculate $\Delta Q$ since the original paper \cite{blondel2008fast} states:}
\begin{quote}
"A similar expression is used in order to evaluate the change of modularity when i is removed from its community. In practice, one therefore evaluates the change of modularity by removing i from its community and then by moving it into a neighbouring community."
\end{quote}
\textit{Surprisingly, this similar expression to compute the change in modularity $\Delta Q$ for the initial removal is never mentioned in the original paper itself \cite{blondel2008fast} ,as well as any other literature we came across during our investigations. We discuss whether or not the formula is incomplete and the implications in \S\ref{sec:formulaComplete?}.} 

Moving on, the algorithm starts by assigning each node of the network to its own community and then iterates through all in Step 1.
The way how it reduces the network in the merging step 2 and the efficient calculation of modularity change are the reasons why the runtime is more efficient than other similar approaches that aim to maximize modularity as well as other community detection methods overall.
Its biggest advantage is that it delivers very good results in terms of modularity in a runtime-complexity of $O(n\cdot \log_{2}n)$ where $n$ is the number of nodes in the network~ \cite{lancichinetti2009community}, which outperforms many other community-detection algorithms, as shown by the original authors of the method \cite{blondel2008fast}.

\subsubsection{Planned Adaptations/Experiments}
\label{LouvainAdapt}
We investigated the following modifications of the Louvain algorithm:
\begin{itemize}

\item First, we used the formula that computes the modularity of the whole current partitioning for the network. This would be computationally inefficient, but it would further help to clarify whether the formula for the change in modularity, as discussed in \S\ref{sec:LouvainDesc} is  complete or not and it would be interesting to see the quality improvements (in terms of modularity) by using this approach.
\item Then, we checked how to improve the community assignment of nodes and save iterations by chaining community assignments together. In its normal form, the algorithm assigns each node to the community of the best neighbours or the assignment stays the same. This step is done for every node. Hence, in this way, there might be some inefficient assignment where, e.g., node 1 is assigned to $c_{2}$, node 2 is assigned to $c_{3}$, and node 3 is assigned to $c_{4}$, and we propose to instantly merge nodes  1,2,3 together into the same community.
%
\item Additionally, we also studied the effect of the node order (the order in which the nodes get iterated through in step 1 of the algorithm). The original paper~\cite{blondel2008fast} states that preliminary results on several test cases seem to indicate that the order of the nodes does not have a significant influence on the result quality in terms of overall modularity score. We therefore randomize the order of nodes in our implementation and want to see how the results differ by running the algorithm repeatedly on the same datasets.
\item Lastly, we attempted to remove some of the greedy aspects of the algorithm by removing the merging step. This of course greatly increases the runtime complexity, but it would be interesting to see if one could make such adaptations to get a better result (in terms of modularity) and in essence trade runtime complexity for result-quality. This would make sense if one would e.g. only use the algorithm on a smaller network/graph where quality would be more important than execution speed.
\end{itemize}

\noindent These adaptations and experiments are discussed in \S\ref{sec:louvainResults}.

\subsection{Fastgreedy}
\label{sec:FG}
\subsubsection{Overview}
\label{sec:FGDesc}
The following algorithm was not given a name by it original authors, and hence we use the name \textit{Fastgreedy}, as proposed by Yang et al.~\cite{YangZhao2016ACAo}. 
The Fastgreedy algorithm was first published by Clauset et al.~\cite{ClausetAaron2004Fcsi} in 2004.
Similar to the Louvain Method discussed in \S\ref{sec:Louvain}, this algorithm is based on modularity.

This is a modification of the algorithm proposed in~\cite{NewmanM.E.J2003Fafd}. Both algorithms use a greedy optimization approach or top-down approach, where at the beginning each vertex represents a community on its own. 
The algorithms repeatedly join two vertices together whose combination produces the largest increase in $Q$. 
In other words, in a network which consists of $n$ vertices, we are left with $n-1$ such joins at the end of the process, at which point the algorithm stops. 
The most straightforward implementation of this algorithm, and also the only one considered in~\cite{NewmanM.E.J2003Fafd}, is to store the adjacency matrix of a given graph as an array of integers, where rows and columns are repeatedly merged as the corresponding communities are merged. 

The Fastgreedy algorithm discussed in this paper achieves to be more efficient in speed and memory usage by eliminating these needless operations. The operation of the algorithm involves finding the changes in $Q$ that would result from the amalgamation of each pair of communities and then choosing the largest of them and performing the corresponding join of communities. 

\subsubsection{Implementation}
\label{sec:FGImpl}
In this enhanced version of the algorithm proposed in \cite{NewmanM.E.J2003Fafd}, rather then maintaining the adjacency matrix and calculating the $\Delta Q_{ij}$, a matrix of updates of $\Delta Q_{ij}$ is maintained. 
The reason is that since joining two communities with no edges between them can never increase $Q$, only the $\Delta Q_{ij}$ need to to be stored for pairs $i, j$ that are connected by one or more edges.
To keep track of the largest $Q$ the authors also make use of of efficient data structures to keep track of the largest $\Delta Q_{ij}$. These improvements result in considerable savings both in time and memory.
There are a total of three data structures that need to be maintained throughout the process:

\begin{enumerate}
    \item A \emph{sparse matrix} containing $\Delta Q_{ij}$ for each pair $i, j$ of communities with at least one edge between them. The authors here propose to store each row of the matrix as an balanced binary tree so that each element can be found or inserted at a time complexity of $O \left(log n \right) $ and as an max-heap, so that the largest element can be found in constant time.
    \item A \emph{max-heap} $H$ containing the largest element of each row of the matrix $\Delta Q_{ij}$ along with the labels $i, j$ if the corresponding pair of communities. 
    \item A \emph{list} or \emph{array} with elements $a_{i}$, where $a_{i} = \frac{k_{i}}{2m}$. $k_{i}$ is the degree of a vertex i and $m$ denotes the number of edges in the given graph.
\end{enumerate}

As described in \S\ref{sec:FGDesc}, in the initial state every vertex represents its own community. 
Now, in the case that vertices $i$ and $j$ are connected, the $\Delta Q_{ij}$ is updated by $\Delta Q_{i,j} = \frac{1}{2m} - \frac{k_{i}k_{j}}{\left(2m\right)^{2}}$.
If the two vertices are not connected, $\Delta Q_{ij} = 0$.
These steps are performed for every vertex $i$ in the network.

\subsubsection{Algorithm Steps}
\label{sec:FGSteps}
The algorithm consists of two steps:

\begin{enumerate}
    \item In the first step the algorithm calculates the initial values of $\Delta Q_{ij}$ and $a_{i}$ according to the update formulas stated above. Then, the max-heap $H$ is populated with the largest element of each row of the matrix $\Delta Q$.
    \item After the initialization of the sparse matrix  $\Delta Q$ with the calculated values in Step 1 and populating the max-heap with the largest values, the largest value $\Delta Q_{ij}$ from $H$ is selected. The communities corresponding to the largest value $i$ and $j$ are joined (note that a community can consist of a single vertex). After joining the corresponding communities, the matrix $\Delta Q$, the max-heap $H$ and array $a$ need to be updated according to these update rules:
    \begin{enumerate}
        \item If the community $k$ is connected to both $i$ and $j$, then the update rule is: $\Delta Q'_{jk} = \Delta Q_{ik} + \Delta Q_{jk}$.
        \item If the community $k$ is connected to $i$, but not $j$, then the update rule is: $\Delta Q'_{jk} = \Delta Q_{ik} - 2a_{j}a_{k}$.
        \item If the community $k$ is connected to $j$, but not $i$, then the update rule is: $\Delta Q'_{jk} = \Delta Q_{jk} - 2a_{i}a_{k}$.s
    \end{enumerate}
    \item Step 2 is repeated until only one community remains. In Step 2, when joining two communities $i$ and $j$, the algorithm labels the resulting community as $j$. Also only the $j$th row and column have to be updated and the $i$th row and column can be removed from the matrix. 
\end{enumerate}

\section{Results \& Discussion}
\label{sec:discussion}

To evaluate, test, and compare the implemented algorithms we mainly used the popular \textit{Karate Club dataset} from Zachary~\cite{zachary1977information} with 77 edges and 34 nodes.
It is considered ``\textit{a standard benchmark in community detection}''~\cite{FortunatoSanto2010Cdig}: each node corresponds to a member of the club, where the edges represent mutual friendship.
Additionally we used small random graphs for testing purposes.
Lastly, to check the runtime of the Louvain algorithm on large network instances, we utilized the \textit{DBLP collaboration network}~\cite{yang2015defining} dataset with \num{317080} nodes and \num{1049866} edges.

We implemented all of these algorithms and their enhancements from scratch in Python 3.7.
NetworkX 2.5 was used to generate the visualizations from the resulting clusterings.

We next present our evaluations and discussion on each of the four selected algorithms, starting with Agglomerative (\S\ref{AggloDiscuss}) and Divisive (\S\ref{subsec:dhc}) Hierarchical Clustering, followed by the Louvain method (\S\ref{sec:louvainResults}) and Fastgreedy (\S\ref{subsec:fg}).

\subsection{Agglomerative Hierarchical Clustering}
\label{AggloDiscuss}
\subsubsection{Experiments \& Results}
 Through a multitude of test runs, we identified interesting and unexpected behavior, which we discuss in this section. 
 Due to space limitations, in this section we restrict ourselves to present selected findings using the Euclidean distance, as the cosine similarity and Pearson correlation coefficient functions as described above did not result in a useful spatial clustering comparable to the other algorithms, and instead clustered nodes in a visually incoherent way. 
 These and other plots, as well as all the discussed code, can be found in~\cite{Repo}.

\begin{figure} 
    \centering
    \subfloat[No self-neighboring \label{1a}]{%
    \includegraphics[width=0.49\linewidth]{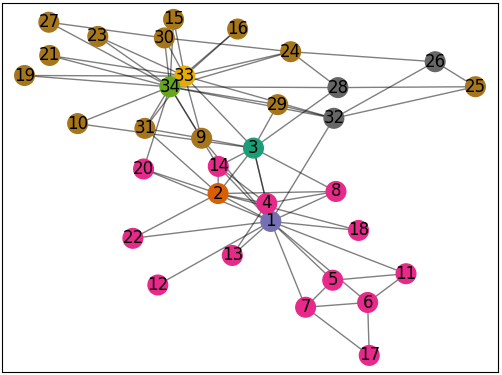}}
    \hfill
  \subfloat[Improved Cluster cohesion through the use of self-neighboring \label{1b}]{%
        \includegraphics[width=0.49\linewidth]{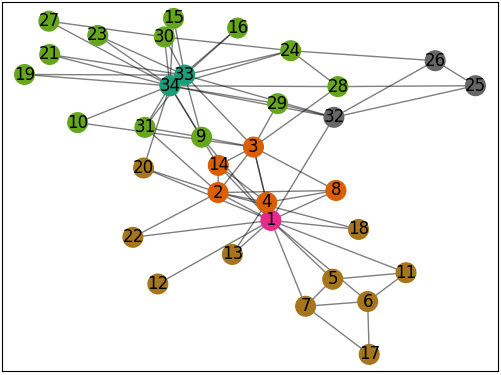}}
    \\
  \caption{Agglomerative Clustering with Euclidean Distance, maximum-linkage and the relative HSL level \textit{rel=0.3} on the Karate Club dataset}
    \label{fig:selfNeighboringComparison} 
    
\end{figure}

Using the \emph{Euclidean distance function}, together with \emph{maximum linkage} and $\textrm{rel}=0.3$ results in a clustering of the Karate Club dataset as shown in Fig.~\ref{fig:selfNeighboringComparison} (a).

This figure also illustrates what agglomerative clustering, using Euclidean distance, identifies first and foremost: well-connected nodes. The same behaviour can still be observed for larger HSL levels. 
This result seems unexpected, as one might anticipate the use of Euclidean distance to result in more tightly grouped node clusters, as it is, for instance, outlined in~\cite{cicada}. 
%
To reveal some of these shortcomings, consider applying the Euclidean distance formula from \S\ref{sec:ACImpl} to the nodes A and B from the three example graphs depicted in Fig.~\ref{fig:3Graphs}: for Graph 1 $d_{AB}=2$, for Graph 2 $d_{AB}=0$, and for Graph 3 $d_{AB}=2$ again. Here, a node does not count itself as its own neighbor, effectively making A and B in Graph 1 further apart than in Graph 2. 

\subsubsection{Adaptation: Self-neighboring}

\begin{figure}[t]
\centering
\includegraphics[width=0.4\textwidth]{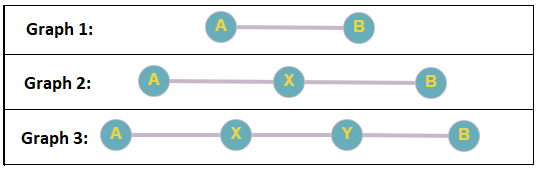}
\caption{Illustration of the intuition for self-neighboring}
\label{fig:3Graphs}
\end{figure}

As such we experimented on how the neighbor matrix $n$ is being constructed, applying an idea we denote as \emph{self-neighboring}. By having each node count itself as a neighbor, we can avoid the above situation and change the distances between A and B in Fig.~\ref{fig:3Graphs} to Graph 1: $d_{AB}=0$; Graph 2: $d_{AB}=2$; Graph 3: $d_{AB}=4$.

After testing this new approach, we observed that, for minimum-linkage, self-neighboring nearly always managed to produce a higher amount of total clusters than the conventional way: even with the same HSL levels, self-neighboring results in more diverse groups, though the overall effect of clustering central and well-connected node first still remained.
For maximum-linkage, the exact opposite was observed: self-neighboring combined with maximum-linkage tends to produce less outliers, often forming more cohesive groups. 
When comparing Fig.~\ref{fig:selfNeighboringComparison} (a) and Fig.~\ref{fig:selfNeighboringComparison} (b), we see how this manifests: with node 2 and 3 being joined with their surrounding cluster, as well as node 25 joining its neighbors (which is especially interesting because, in Fig.~\ref{fig:selfNeighboringComparison} (a), node 25 shared a cluster with \textit{none} of its neighbors) and node 33 and 34 now belonging to the same group as well, it can be observed how self-neighboring paired with maximum linkage increases cohesiveness and shrinks the overall amount of clusters, without needing to raise the HSL level. 
A similar effect also occurs for average linkage.

Based on these results, we propose \emph{self-neighboring}, paired with maximum linkage, as a possible approach to computing neighbor matrices when trying to group clusters based on their Euclidean distance, especially if it is desired to reduce outliers.

\subsection{Divisive Hierarchical Clustering (Girvan-Newman)}
\label{subsec:dhc}
\subsubsection{Results, adaptions and experiments}
\label{sec:girvanNewmanResults}

\begin{figure}[t]
\centering
\centering
\includegraphics[width=0.42\textwidth]{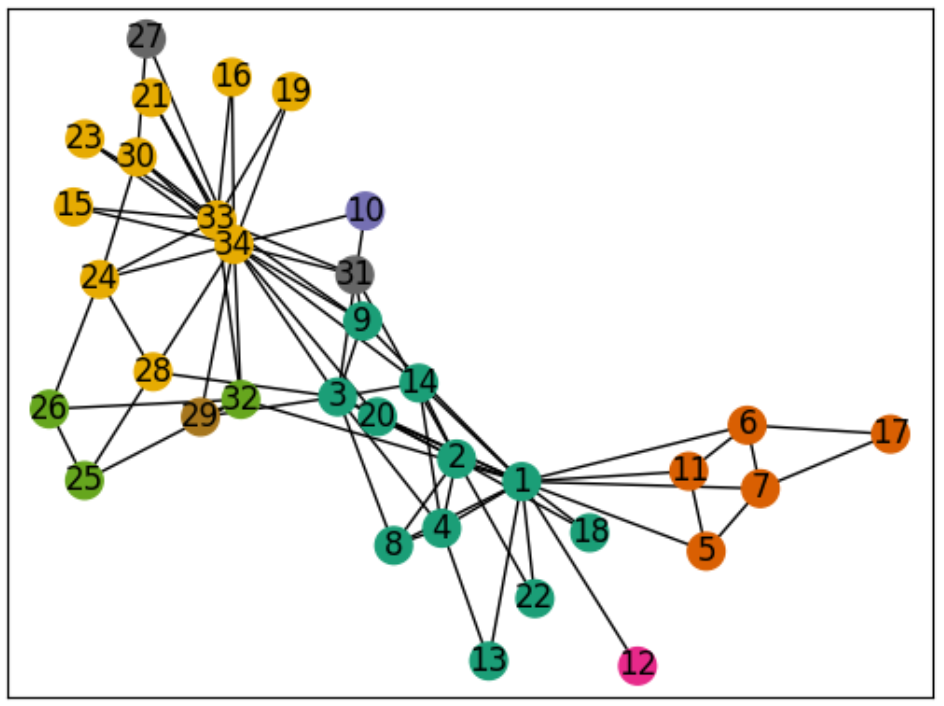}
\caption{Karate Club dataset clustered into 8 communities by our implementation of the Girvan-Newman algorithm}
\label{fig:GirvanNewmanFigure}
\end{figure}

We start this algorithm section with an illustration, where Fig.~\ref{fig:GirvanNewmanFigure} shows the result of clustering the Karate Club dataset into 8 communities.
While one can expect good results in general, one of the biggest drawbacks of the Girvan-Newman algorithm is its computational performance. The breadth-first search has to run on each node of the graph and has to be repeated after each edge removal, hence the algorithm has an overall time complexity of $O(EN(E+N))$.

As such we experimented with skipping the repetition of the edge betweenness calculation as proposed by Meghanathan~\cite{meghanathan2016NOVER}.
Hereby, the scores for each edge are calculated at the beginning of the algorithm and stay unchanged during the runtime of the program. 
The edges are cut in decreasing order of their initial betweenness scores. 
Hence the time complexity of this method is only $O(N+E)^2$ and as thus our results show a good boost in performance. 
However, as expected, community detection tends to be not as precise as with the original iterative approach. 

\subsection{Louvain Method}
\label{sec:louvainResults}
\subsubsection{Reducing Greediness for Quality}
\label{sec:lessGreed}
In order to adapt the Louvain algorithm, we decrease its greedyness by removing the merging step (2) of the algorithm, which reduces the graph by creating one big node out of all nodes belonging to the same community. 
Additionally, we also used the total modularity formula for the whole graph (for reasons see Section \S\ref{sec:formulaComplete?}), which on the other hand increased the runtime complexity.
Herein, the general idea was to be able to trade higher runtime complexity for potentially higher quality results.

On our small test graphs,  this led to the same ``optimal'' result that we achieved using the standard greedy version, as well as a standard (merging) version that used the total modularity formula. 
However, this result was achieved independently of the node order, and hence there is a qualitative benefit,
%
as the greedy method had some node orderings that led to a slightly worse results, since the assignment of individual nodes was sometimes different.

In the bigger graphs, maybe surprisingly, the optimal result also remained the same. 
However, in general, the results tend vary a lot more and there is a trend for more smaller communities. 
Moreover, there are also potential results that are worse than any of the results of the standard/\textit{normal} version. 
We visualize these findings in the violin plot in Fig.~\ref{fig:LouvainViolin}.
If one considers the violins of \textit{normal} and \textit{total}, the same algorithm that just uses the total modularity formula instead of the one calculating local change,\footnote{For reasons why the results differ from \textit{normal} we refer to \S\ref{sec:formulaComplete?}.} and compares them to the violins of \textit{noMerge} and \textit{totalNoMerge} to clearly see that the first 2 versions in general yield better results than their non-merging counterparts. 
A summary of the used methods is also depicted in Table~\ref{tab:LouvainResults}.
Therefore, we can state that the merging step does not only contribute to a drastic reduction of runtime complexity, but also helps in creating a better partitioning.

\subsubsection{Completeness of modularity change formula}
\label{sec:formulaComplete?}
At first glance, the standard version of the Louvain-algorithm as described in \S\ref{sec:LouvainDesc} seems complete, and our results were appropriate and matched the results of the NetworkX Python library, as well as manual calculations using the total modularity formula on a small test graph.


However, while implementing the less greedy version of the algorithm (see \S\ref{sec:lessGreed} above), we encountered the problem that the algorithms would not terminate and loop, indefinitely swapping node community assignments.

The first check performed was on ties in community assignment, where having a node in one community or the other would always lead to the same small increase and the algorithm therefore always swaps them around. This turned out to not be the case,
as even changing the implementation in a way that such assignments would not take place did not solve the issue. 

The real reason for that behaviour was that the formula, which as stated by the original authors \cite{blondel2008fast} and briefly touched upon in \S\ref{sec:LouvainDesc}, is only partial and only captures the increase/decrease in modularity one gets when assigning the node to another community, but not how the removal of the node from its current community affects it. 
We obtained further validation by calculations using the ``incomplete'' formula on a small graph where:
\begin{itemize}
    \item the Assignment of a node to the same community would still result in a modularity increase $\rightarrow$  this should not change modularity at all,
    \item the removal of a node from a bigger community and assignment to a new community consisting only of that one node resulted in a modularity change of "0" $\rightarrow$ this should decrease modularity.
\end{itemize}

\begin{figure}[t]
\centering
\includegraphics[width=0.4\textwidth]{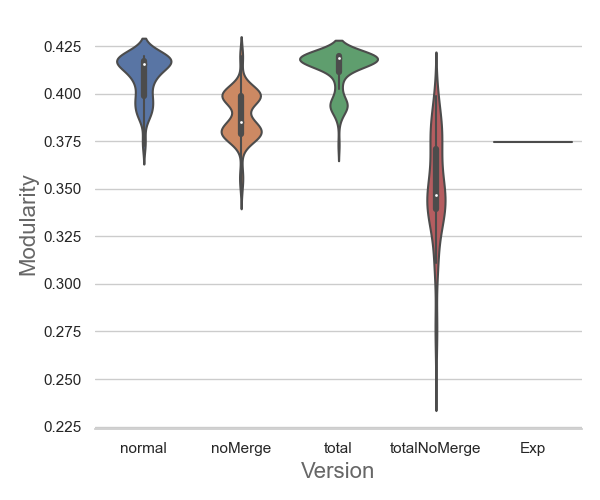}
\caption{Modularity scores for various Louvain-Algorithm versions (100 runs per version) on the Karate Club dataset~\cite{zachary1977information}}
\label{fig:LouvainViolin}
\end{figure}

Another evidence for the incompleteness of the formula is the violin plot in Fig.~\ref{fig:LouvainViolin}, where one can see that the version using the total modularity formula (\textit{total}) led to better results more often than the \textit{normal} one---if the formula would be complete, then the behaviour of the 2 versions should be exactly the same.

We therefore believe that the formula that is found in the original paper~\cite{blondel2008fast} and other sources is not complete. Surprisingly, the second part that calculates the cost of removal from the current community seems to be missing as well.

Notwithstanding, the algorithm works well without this part of the formula (in the best case we reach the same modularity score on the Karate Club dataset~\cite{zachary1977information} as the original authors~\cite{blondel2008fast}),  since the cost of removal seems to be often insignificant when working with the standard version of the algorithm (especially after nodes get merged in Step 2). 

Interestingly enough, as also depicted by the violin plot (Fig.~\ref{fig:LouvainViolin}, label \textit{totalNoMerge}), the total and complete formula performs worse than the ``incomplete'' one if we omit the merging step as described in \S\ref{sec:lessGreed}. 
The results vary the most and in 100 runs, the best possible result was not achieved.
We cannot pinpoint a definitive explanation for this behavior---our assumption is that even though the ``incomplete'' formula for just the change calculation sometimes assigns nodes to initially non-optimal communities, this might lead to more favorable assignments later on.\footnote{Similar to evolutionary algorithms, were sometimes one needs to go into a seemingly unfavorable direction to escape local maxima and move towards the global maximum.}

\subsubsection{Faster merging experiment}
\label{sec:fasterMerging}
We also adapted the algorithm in by how nodes are assigned to clusters in a way that assignments get merged/combined.
If, e.g., node 1 gets assigned to $c_{2}$, node 2 gets assigned to $c_{3}$, and node 3 gets assigned to $c_{4}$, they are put all in the same cluster/community as their assignments are merged.

\begin{figure} 
    \centering
  \subfloat[Best Partitioning using our standard Version of the Louvain-Algorithm on the Karate Club dataset \label{1a}]{%
       \includegraphics[width=0.49\linewidth]{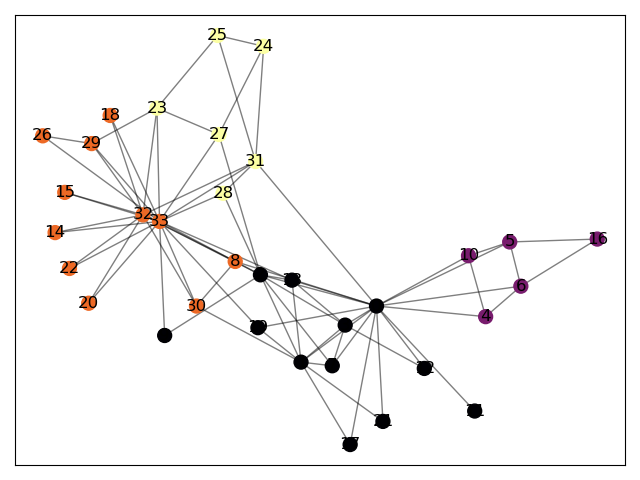}}
    \hfill
  \subfloat[Partitioning using our experimental Version of the Louvain-Algorithm on the Karate Club dataset \label{1b}]{%
        \includegraphics[width=0.49\linewidth]{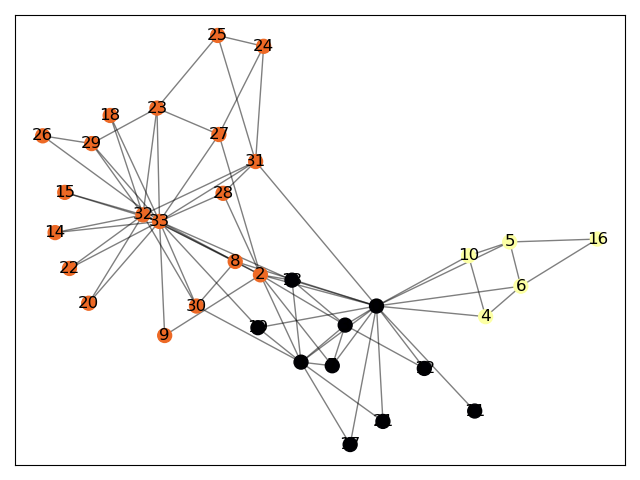}}
    \\
  \caption{Partitionings using (a) the standard Louvain-Version and (b) our experimental Louvain-Version}
  \label{fig:LouvainBestNormalExp} 
\end{figure}

This led to a deterministic algorithm version, where the results are always the same, as visible in Fig.~\ref{fig:LouvainViolin} with the label \textit{Exp}. 
Since the assignments are only done at the end of each iteration, they no longer are dependant on the order in which one iterates through the nodes.
The result had a worse modularity score than all or most of the partitionings that were determined by the standard as well as the other version.
In general, the results also had fewer clusters than the near optimal partitioning in terms of modularity.
However, when considering the resulting community assignments, as in Fig.~\ref{fig:LouvainBestNormalExp} (b), the results are still good enough.
This combined with a slightly faster run-time (see Table~\ref{tab:LouvainResults} and the advantage of deterministic results 
might make it attractive in certain situations.

\subsubsection{Runtime complexity}
Due to the inconsistencies found w.r.t.\ to the described and expected behavior, we also performed a small validation of the practical computational complexity.
We can confirm the efficiency and low runtime of the Louvain algorithm by using our standard version of it on the DBLP dataset~\cite{yang2015defining} with \num{317080} nodes and \num{1049866} edges, where it terminates in roughly 6 minutes with a modularity score of 0.805. 
This result, While still being significantly higher than the 3 seconds that the original authors state~\cite{blondel2008fast} in their paper, for a dataset of comparable size (approximately $325$K nodes and 1 million edges)%
, still points towards a good runtime complexity and we hence believe that the longer execution time we observed can be explained by using weaker hardware and a not as efficient implementation and Python technology stack. 

\subsection{Fastgreedy}
\label{subsec:fg}
\subsubsection{Results, Adaptations and Experiments}
\label{FGImprove}
\begin{figure} 
    \centering
  \subfloat[Six cluster remaining\label{1a}]{%
     \includegraphics[width=0.5\linewidth]{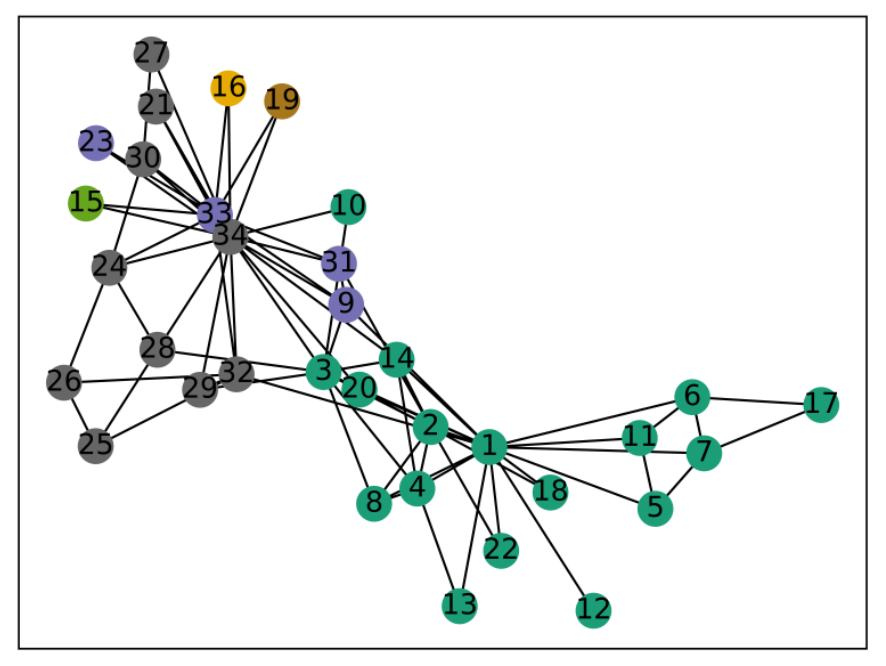}}
    \hfill
  \subfloat[Three cluster remaining\label{2b}]{%
      \includegraphics[width=0.5\linewidth]{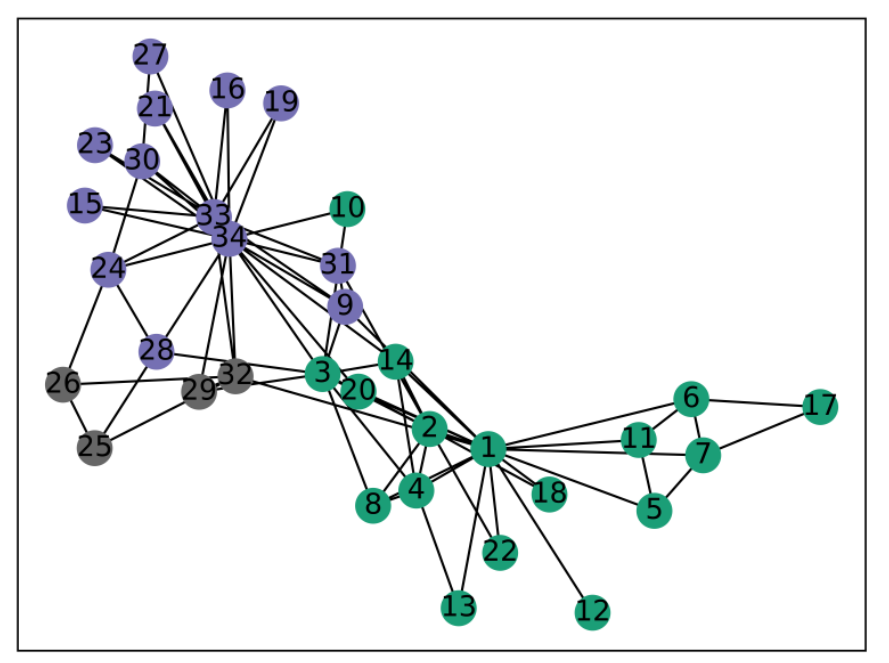}}
    \\
  \caption{Two clustering steps by the Fastgreedy algorithm}
  \label{fig:FGSnapshots} 
\end{figure}

Fig.~\ref{fig:FGSnapshots} shows two snapshots of the community clustering process on the Karate Club graph \cite{zachary1977information} as it was produced by our final Fastgreedy implementation.
%
%
%
%
Regarding possible adaptations, Fastgreedy does not leave much room for improvements compared to the other algorithms, such as changing the distance functions to obtain better results concerning community clustering, and we already examined which adaptations could be made to another modularity based and greedy algorithm in \S\ref{LouvainAdapt}. 

However, we performed some tweaks on the implementation part of the algorithm, which we describe next.
The authors used a balanced binary tree to store each row of the matrix $\Delta Q$.
Since there is no balanced binary tree module in the stdlib of Python we chose to use simple Python dictionaries instead. Dictionaries are implemented as hash tables, which have a lookup and insertion time of $O(1)$. 
Instead of creating a single balanced binary tree for each row, we create a single dictionary for all rows. 
Every entry in the dictionary represents a row of the matrix $\Delta Q$, and within an entry we created, again, a dictionary for each row element with its corresponding value $\Delta Q_{ij}$.
This approach may take up more memory, but is more efficient in looking up the values.
Hence, it might be beneficial, as the memory capacity of modern computing systems has vastly expanded since the original paper publication in 2005.

Next, to perform further experiments, we included operations on a real sparse matrix. 
In the author's approach, the matrix is represented only by balanced binary trees for each row. 
Python's library \textit{scipy} nowadays offers a wide variety of sparse matrices, such as \textit{csc}, a compressed sparse column matrix or \textit{csr}, a compressed sparse row matrix. Since our operations concern mostly rows we went along with the \textit{csr} matrix. Using a matrix brings the advantage that we can search and insert values $i$ and $j$ in $\Delta Q_{ij}$ only by using the matrix indices, which results in a runtime complexity of $O(1)$. However, using a real matrix representation brings some downsides as well. 
First deleting rows and columns, as it was stated by the authors in the original paper, is not longer possible, since removing them would result in wrong indexing. 
Therefore rows and columns that are no longer of use need to be kept, which results again in more memory usage. 
As stated earlier, we believe this trade-off to be worthwhile.
Further evaluations could be conducted to decide which of the two approaches uses less memory.
In terms of the overall runtime, the results should be similar, as both methods use a lookup and insertion time of only $O(1)$. 

The last improvement we did implement was suggested by the authors themselves. They proposed that rather of implementing a heap $H$ for each row, as it was also the case with the balanced binary tree, it would be more feasible to implement a larger heap for the whole matrix $\Delta Q$. This approach not only simplifies the implementation, but, according to the authors, the average case running time is better compared to the multiple heap version.
Our implementation is available at~\cite{Repo}.

\newcolumntype{Y}{>{\raggedright\arraybackslash}X} 
\begin{table}[t]
  \caption{Comparison of Louvain versions (100 runs, Karate Club dataset, Intel i7-3770k CPU@3.5GHz, 16GB RAM)}
  \label{tab:LouvainResults}
  \begin{tabular}{ | m{0.95cm} | m{3.2cm}| m{0.79cm} | m{0.79cm} | m{0.85cm} | }
  \hline
    \textbf{Label} & \textbf{Description} & \textbf{Max. Score} & \textbf{Min Score} & \textbf{Avg. runtime (ms)}\\
  \hline
    normal & Algorithm as described in the original paper \cite{blondel2008fast}. & 0.41979 & 0.38108 & 3.899 \\ 
      \hline
    total & Algorithm as described in the original paper \cite{blondel2008fast} that uses the formula for the total modularity of the whole graph to calculate change instead of the formula that directly calculates the change. & 0.41979 & 0.3792 & 440.799\\
   \hline
    noMerge & like \textit{"normal"} but the merging step (2.) of the algorithm (see \S\ref{sec:LouvainDesc} is omitted and the algorithm always works with the full set of initial nodes. & 0.41979 & 0.35396 & 57.314\\
      \hline
    total
    NoMerge & like \textit{"total"} but the merging step (2.) of the algorithm (see \S\ref{sec:LouvainDesc} is omitted and the algorithm always works with the full set of initial nodes. & 0.39981 & 0.24499 & 1966.91  \\ 
   \hline
    Exp & Experimental version described in \S\ref{sec:fasterMerging}, creates less bigger clusters and is deterministic. & 0.37443 & 0.37443 & 3.349\\
    
 
  \bottomrule
\end{tabular}
\end{table}




\section{Conclusion and Outlook}\label{sec:conclusion}
\vspace{-1mm}
In this paper, we not only explored and implemented four common community detection algorithms from the ground up, in order to perform a technical investigation w.r.t.\ their algorithmic behaviour \emph{on paper} versus \emph{in code}, but also proposed and evaluated a range of enhancements based on our findings.
We investigated the partitioning behavior of Agglomerative Clustering and how it is affected by the choice of distance function, linkage function, and HSL level. Based on our results, we proposed \emph{self-neighboring} as an alternative way to construct the neighbor matrix, which manages to improve cohesion by \emph{closing gaps} between clusters.
The second algorithm we studied is the well-known Girvan-Newman algorithm~\cite{girvanNewman}. 
In order to improve its relatively slow runtime,  we \emph{validated} the proposed adaption by Meghanathan \cite{meghanathan2016NOVER} and can confirm a boost in performance.
For the Fastgreedy \cite{ClausetAaron2004Fcsi} algorithm, we performed a fine-tuning, where we modified some of the initially proposed data structures. 
We used a compressed sparse row matrix for \emph{efficient row slicing} and also implemented the proposal of the authors to use one max-heap for all matrix rows. Moreover, we used dictionaries instead of binary search trees, resulting in more efficient operations. 
Regarding the Louvain Algorithm, we discovered that the full formula for modularity change was not disclosed in the original paper. We as thus investigated and analyzed the impact of using the partial formula by comparing the results generated by its use with the ones created by the total modularity formula. Additionally, we also experimented with the merging step of the algorithm as we tried to omit it and see how it affects the results.
Our main results in this context was that the merging step of the algorithm not only \emph{reduces the runtime} of the algorithm but also helps to create \emph{better} results in terms of \emph{modularity}, but that at the same time, the incomplete formula still works well---except when omitting the merging step, but can nonetheless cause an indefinite loop in certain situations.
We moreover proposed an adaptation, where we changed the method how communities are assigned, to the Louvain Method, that is faster, deterministic, and creates bigger clusters compared to the normal version.


\vspace{-1mm}
\subsection{Future Work}
\vspace{-1mm}
\label{sec:futureWork}

One possible direction for future work would be to investigate and implement more clustering algorithms with other underlying paradigms (such as, e.g., using random walks as presented by Pons et al.~\cite{pons2005computing}).

Another possible direction would be to investigate the already implemented algorithms on further datasets. Especially with the agglomerative clustering approach, where one can arbitrarily exchange the distance function, there is good potential for further exploration and investigation by leveraging more intricate distance functions.

Regarding the Louvain method, a further step could be to derive the complete formula to calculate the change (as discussed in \S\ref{sec:formulaComplete?}) and repeat the experiments as well as try to optimize the performance, as there seems to be room for improvement, when comparing it to the speed that the authors of the original paper~\cite{blondel2008fast} observed. Furthermore, there have also been orthogonal adaptations to the Louvain method, such as, e.g., by De Meo et al.~\cite{de2011generalized}, who were able to slightly improve the results of the original method using a novel measure of edge centrality based on \textit{k-paths}: this measure could also be of interest for the Girvan-Newman \cite{girvanNewman} algorithm.

\noindent\textbf{Reproducibility}
In order to guarantee reproducibility and facilitate other researchers to build upon our work, we make our source code publicly available~\cite{Repo}.

\vspace{-2mm}




{\footnotesize
\bibliographystyle{IEEEtran}
\bibliography{references}
}
\end{document}